\documentclass{iaus}
\usepackage{graphicx}

\title[Magnetic fields in the Herbig Ae/Be stars] 
{Discovery of fossil magnetic fields in the intermediate-mass pre-main sequence stars}

\author[E. Alecian ]   
{Evelyne Alecian$^{1,2}$%
Gregg A. Wade$^1$
Claude Catala$^2$
}

\affiliation{
 $^1$Royal Military College of Canada, PO Box 17000, Stn Forces, Kingston K7K 7B4, Canada \\
 $^2$Observatoire de Paris, LESIA, place Jules Janssen, 92190 Meudon, France \\
 }

\pubyear{2009}
\volume{259}  
\pagerange{119--126}
\date{?? and in revised form ??}
\jname{Cosmic Magnetic Fields: from Planets, to Stars and Galaxies}
\editors{A.C. Editor, B.D. Editor \& C.E. Editor, eds.}
\begin{document}

\maketitle

\begin{abstract}
It is now well-known that the surface magnetic fields observed in cool, lower-mass stars on the main sequence (MS) are generated by dynamos operating in their convective envelopes. However, higher-mass stars (above 1.5 M$_{\odot}$) pass their MS lives with a small convective core and a largely radiative envelope. Remarkably, notwithstanding the absence of energetically-important envelope convection, we observe very strong (from 300 G to 30 kG) and organised (mainly dipolar) magnetic fields in a few percent of the A and B-type stars on the MS, the origin of which is not well understood. In this poster we propose that these magnetic fields could be of fossil origin, and we present very strong observational results in favour of this proposal.
\end{abstract}

\firstsection 
\section{Introduction}
The fossil field theory assumes that the magnetic fields observed in the main sequence (MS) A/B stars are remnants from the galactic fields observed among the molecular clouds. Magnetic fields are observed in the molecular clouds and in (MS) intermediate mass stars. Until recently no magnetic fields has been observed in intermediate stages of stellar formation, especially during the pre-main sequence (PMS) phase (except in HD 104237, Donati et al. 1997). We therefore tried to find observational evidence that some PMS intermediate mass stars, the so-called Herbig Ae/Be stars (HAeBe), possess magnetic fields.

\begin{table}
\begin{center}
\caption{Results of the fitting procedures for 4 magnetic HAeBe stars. References: 1: Alecian et al. 2008a, 2: Folsom et al. 2008, 3: Alecian et al., in prep., 4: Catala et al. 2007}
\label{tab}
\begin{tabular}{@{}ll@{}ccccc@{}cl@{}}
\hline
Star & Sp.T. & $P$ (d) & $i$ ($^{\circ}$) & $\beta$ ($^{\circ}$) & $B_{\rm d}$ (kG) & $d_{\rm dip}$ ($R_{*}$) & $B_{\rm dZ}$ (kG) & Ref. \\
\hline
HD 200775 & B3 & $4.3281\pm0.0010$ & $60\pm11$ & $125\pm8$ & $1.00\pm0.15$ & $0.05\pm0.04$ & 3.6 & 1 \\
HD 72106 & A0 & $0.63995\pm0.00009$ & $24\pm10$ & $57\pm5$ & $1.25\pm0.08$ & 0 & 1.25 & 2 \\
V380 Ori & A2 & $4.31\pm0.08$ & $154\pm30$ & $110\pm30$ & $2.6\pm1.5$ & 0 & 4.5 & 3 \\
HD 190073 & A2 & & & & $>0.3$ & & $>1.2$ & 4 \\
\hline
\end{tabular}
\end{center}
\end{table}

\section{Observations and data reduction}
We used the high-resolution spectropolarimeters ESPaDOnS installed on the Canada-France-Hawaii Telescop(CFHT, Hawaii) and Narval installes on the Bernard Lyot Telescope(TBL, Pic du Midi, France). We obtained Stokes $I$ and $V$ spectra of 65000 spectral resolution for a sample of 130 HAeBe stars in the field and in three very young clusters: NGC 2264 (age$\sim$2.6 Myr), NGC 2244 (age$\sim$2.3 Myr), and NGC 6611 (age$<$1 Myr).

We used the Libre ESpRIT reduction package, and we applied the Least Square Deconvolution Method (LSD, Donati et al. 1997), using ATLAS9 masks of appropriate effective temperature and surface gravity, excluding hydrogen Balmer lines and lines contaminated by emission. Figure \ref{fig} show resulting LSD $I$ and $V$ profiles for one of these stars, HD 200775, in which we detect a strong dipolar magnetic field.

\section{Results}

\begin{figure}
\includegraphics[width=4cm]{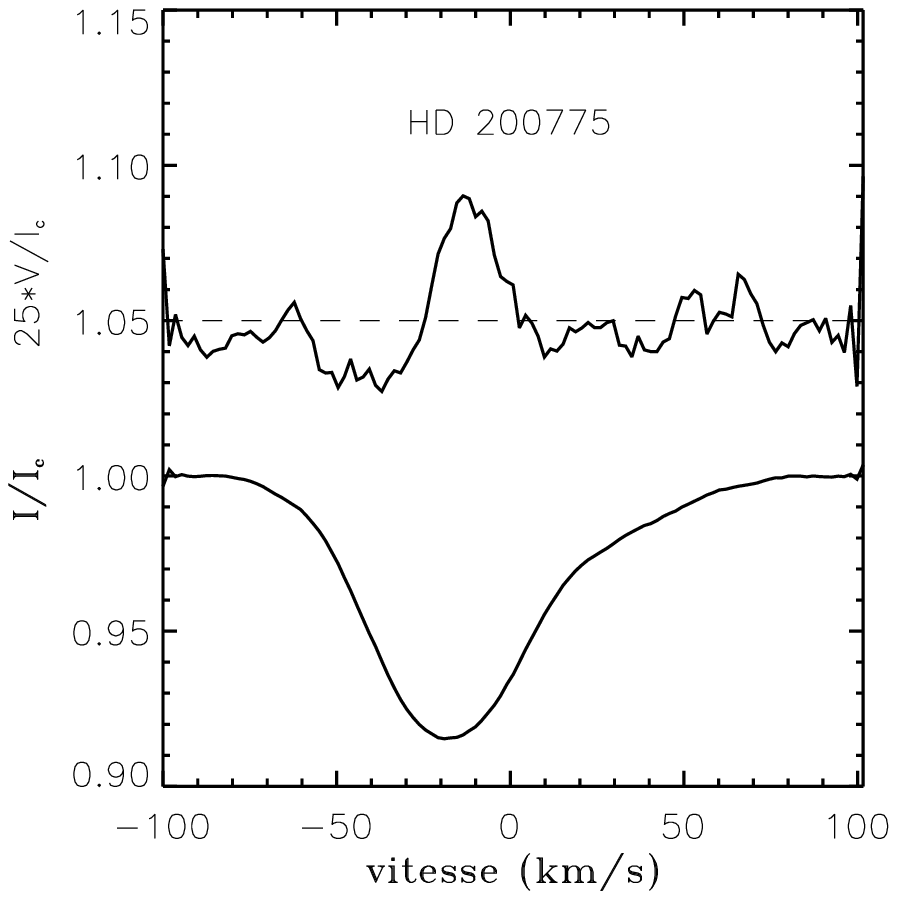}
\hfill
\includegraphics[angle=90,width=8cm]{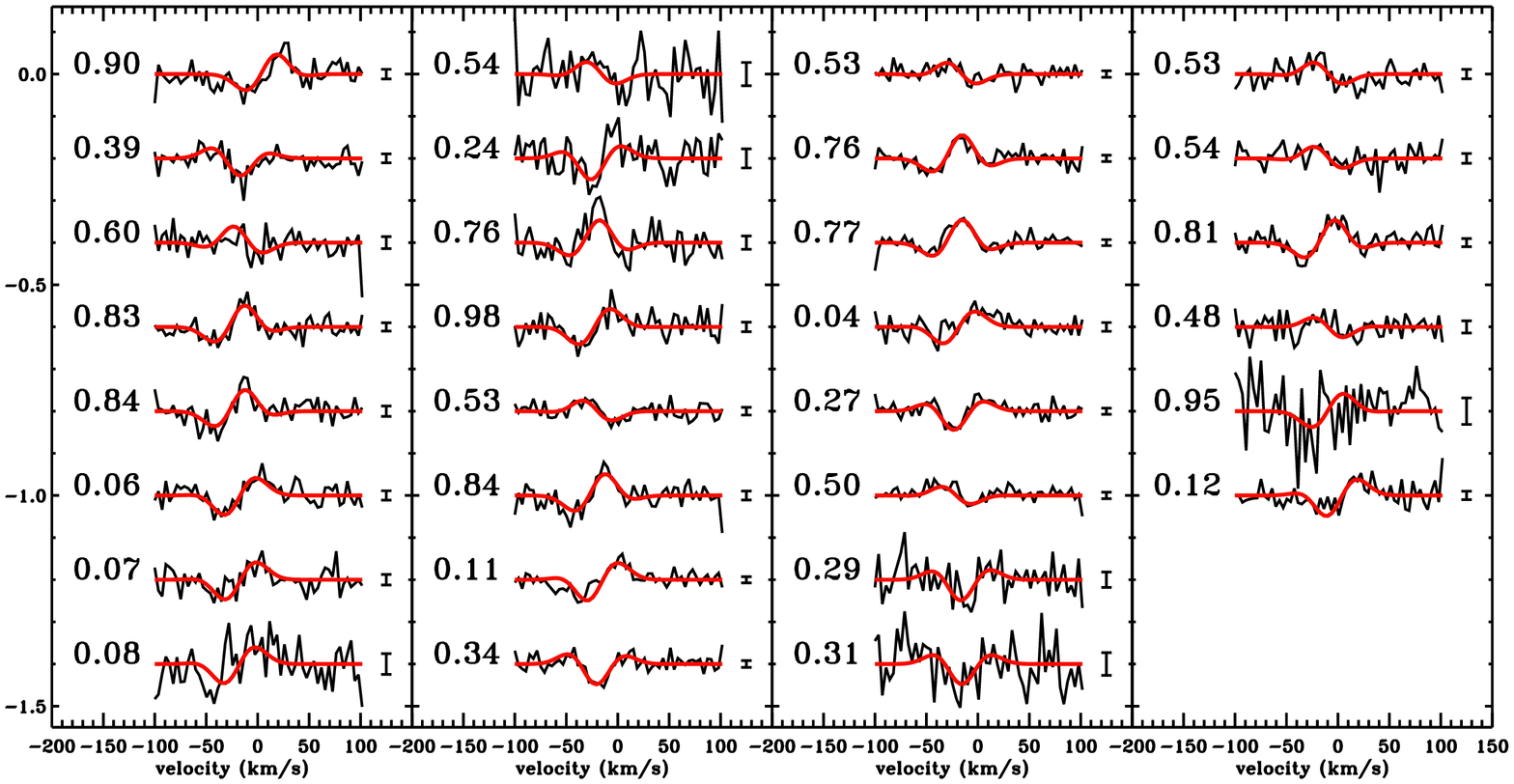}
\caption{{\bf Left:} Stokes $I$ (bottom) and $V$ (up) LSD profils of HD 200775. {\bf Right:} Observed (noisy black) and modelled (red smooth) Stokes $V$ profiles of HD 200775.}
\label{fig}
\end{figure}

We have detected magnetic fields in 7 HAeBe stars (Wade et al 2005, Catala et al. 2007, Alecian et al. 2008ab). For 4 of them, we have characterised their magnetic fields as follows. We monitored these stars in Stokes $V$ to densely sample their rotation cycles. Then we used the oblique rotator model, as described in Stift (1975), to reproduce the variations of their Stokes V profiles (see also Alecian et al. 2008a). The free parameters are: the rotation period ($P$), the reference time ($T_{\rm 0}$), the rotation inclination ($i$), the magnetic obliquity ($\beta$), the dipole strength ($B_{\rm d}$), and the dipole shift ($d_{\rm dip}$) from the center of the star expressed in stellar radii ($R_{*}$). We performed a $\chi^2$ minimisation in order to reproduce our observations. The Fig. \ref{fig} shows the result of this fitting procedure for one star, while Table \ref{tab} gives the values of the fitted parameters.

\section{Conclusion}

We find that around 5\% of HAeBe stars are magnetic, as predicted from the fossil field theory. These stars have strong and large scale magnetic fields stable over more than 3 years, as observed in the MS A/B stars. Finally, assuming the conservation of magnetic flux during the PMS phase, we find that when the magnetic HAeBe stars will reach the ZAMS their magnetic strengths will be between 1.2 to 3.6 kG, similar to the magnetic MS A/B stars. Therefore the magnetic fields of the A/B stars is very likely fossil.

\end{document}